# Highly Anisotropic Excitons and Multiple Phonon Bound States in a Van der Waals Antiferromagnetic Insulator

**Authors:** Kyle Hwangbo[1†], Qi Zhang[1†], Qianni Jiang[1], Yong Wang[2], Jordan Fonseca[1], Chong Wang[3], Geoffrey M. Diederich[1], Daniel R. Gamelin[4], Di Xiao[3], Jiun-Haw Chu[1], Wang Yao[5], Xiaodong Xu[1,6*]

[1]Department of Physics, University of Washington, Seattle, Washington 98195, USA
[2]School of Physics, Nankai University, Tianjin 300071, China
[3]Department of Physics, Carnegie Mellon University, Pittsburgh, Pennsylvania 15213, USA
[4]Department of Chemistry, University of Washington, Seattle, Washington 98195, USA
[5]Department of Physics and Center of Theoretical and Computational Physics, University of Hong Kong, Hong Kong, China
[6]Department of Materials Science and Engineering, University of Washington, Seattle, Washington 98195, USA

[†]These authors contributed equally to this work.

[*]Correspondence to: xuxd@uw.edu

**Abstract:** Two-dimensional semiconducting systems, such as quantum wells and transition metal dichalcogenides, are the foundations to investigate low dimensional light-matter interactions[1,2]. To date, the study of elementary photoexcitation, namely the exciton, in 2D semiconductors with intrinsic magnetic order remains a challenge due to the lack of suitable material platforms[3,4]. Here, we report an observation of excitons coupled to zigzag antiferromagnetic order in the layered antiferromagnetic insulator $NiPS_3$ using both photoluminescence and optical reflection spectroscopy. The exciton exhibits a linewidth as narrow as ~350 μeV with near unity linear polarization in the photoluminescence spectrum. As the thicknesses of samples is reduced from five layers to bilayers, the photoluminescence intensity is drastically suppressed and eventually vanishes in monolayers, consistent with the calculated bandgap being highly indirect for both bilayer and monolayer[5]. Furthermore, we observed strong linear dichroism over a broad spectra range, which shares the same optical anisotropy axis, being locked to the zigzag direction, as the exciton photoluminescence. Both linear dichroism and the degree of linear polarization in the exciton photoluminescence decrease as the temperature increases and become negligible above the Néel temperature. These observations suggest both optical quantities are probes of the symmetry breaking magnetic order parameter. In addition, a sharp resonance in the linear dichroism spectrum is observed with an energy near the exciton photoluminescence. There exist over ten exciton-$A_{1g}$ phonon bound states on its high energy side, which likely result from the strong modulation of the ligand-to-metal charge transfer energy by strong electron-lattice interactions. Our work establishes $NiPS_3$ as a new 2D platform for exploring magneto-exciton physics with strong correlations, as well as a building block for 2D heterostructures for engineering physical phenomena with time reversal symmetry breaking.

**Main text**

Semiconductors with long-range magnetic order have long been pursued as a platform to explore both fundamental magnetism and spintronic applications[6-10]. The elementary photoexcitations in semiconductors are excitons, which are Coulomb-bound electron-hole pairs. Excitons coupled to magnetic order have been studied in several intrinsic magnetic insulators. For instance, in bulk crystals with Néel antiferromagnetic ordering, Davydov splitting resulting from the coupling of localized excitons in adjacent magnetic ions has been extensively studied[11-14]. Coupling of the electronic structure to magnetic order has also led to strong exciton-magnon coupling[10,14] and drastic tunability of excitonic photoluminescence by a magnetic field[13]. In quantum well structures, dilute magnetic semiconductors have served as a testbed for exploring magneto exciton physics. Due to their low spin densities, however, these materials typically show weak inter-ion magnetic interactions and hence uncorrelated spins, posing fundamental limitations for studying physics and applications linked to intrinsic magnetic order[15].

The recent emergence of two-dimensional van der Waals (vdW) magnets may provide a new quantum well platform with intrinsic magnetic order [16,17]. Indeed, highly localized excitons with spontaneous circularly polarized photoluminescence have been observed in ferromagnetic monolayer $CrI_3$ and $CrBr_3$[4,18]. The first principles calculations for $CrI_3$ suggest the ground state exciton is dark and Frenkel type, while several high energy features observed in absorption are charge transfer excitons with Wannier type character[19]. Despite this exciting progress, several challenges remain for exploring features unique to 2D materials. The Frenkel character of the ground state exciton makes it insensitive to layer thickness[4], and thus not suitable for vdW engineering of exciton properties (e.g. moiré excitons). The photoluminescence linewidth of the ground state exciton is broad (~150 meV), while narrow linewidth is desirable for magneto optical physics. Lastly, there is little known about excitons in 2D antiferromagnets with Néel and zigzag type. There is no net magnetic moment in antiferromagnets, and the strong exchange interactions often require high magnetic field to tune the magnetic order. Therefore, it is unknown how excitons are coupled to these antiferromagnetic orders at the atomically thin limit. We anticipate that overcoming these challenges would enable scientists and engineers to harness the features unique to 2D materials to explore fundamental phenomena such as quantum many-body effects of magneto-excitons, provide new means to probe magnetism, and create opportunities for optically controlled spintronic devices based on vdW heterostructures.

Here, we report the observation of 2D excitons in atomically thin, antiferromagnetic $NiPS_3$ which exhibits strong coupling between electronic, magnetic, and lattice degrees of freedom. $NiPS_3$ belongs to a class of transition metal phosphorous trichalcogenides ($APX_3$, A: Fe, Mn, Ni and X: S, Se), which are van der Waals antiferromagnetic insulators[20-24]. Within individual $NiPS_3$ layers, Ni is arranged in a honeycomb lattice structure. The spins are aligned in the zigzag direction along the a-axis, while the adjacent spin chains are *anti*-aligned (Fig. 1a), forming a zigzag antiferromagnetic order. Neutron scattering suggests that the spins are mostly aligned in-plane with only a slight tilt out of plane[24,25]. The interlayer coupling between adjacent layers is ferromagnetic (Fig. 1a). Based on Raman scattering measurements, long-range magnetic order remains down to the bilayer, but is suppressed in monolayers[26]. X-ray absorption studies on bulk crystal samples suggest that NiPS3 is a charge-transfer antiferromagnetic insulator that exhibits strong correlation effects[27].

We perform both photoluminescence and optical reflection spectroscopy measurements on NiPS3 samples with varying thicknesses. All the samples presented in the main text are prepared by mechanical exfoliation of bulk crystals onto 90 nm SiO2/Si substrates in an Argon gas glove box to minimize potential sample degradation in ambient conditions. Samples with thicknesses ranging from several hundred layers down to a single layer are studied. Figure 1b is an optical microscope image of a representative sample with the layer numbers indicated in Figure 1c. The layer numbers were identified through optical contrast and atomic force microscopy imaging.

We first present linear polarization resolved photoluminescence measurements of a thin bulk crystal excited with a 633 nm excitation laser. We define the horizontal and vertical axes according to polarization resolved Raman spectroscopy (Extended Data Figure 1) [26,28]. Figure 1d shows the data at selected temperatures. Notably, at 15K, a pronounced peak with a Lorentzian lineshape is observed at 840.5 nm (1.4753 eV); this is a signature of exciton photoluminescence. The peak width, measured via the full width at half maximum (FWHM), is as narrow as 0.20 nm (350 µeV) (Fig. 1d). This is much narrower than the peak widths measured in 2D transition metal dichalcogenides encapsulated in hexagonal boron nitride[29,30]. Although the exact band edge of NiPS3 and the exciton binding energy are unknown, the optical bandgap has been identified at 1.8 eV[5,27]. We further performed time-resolved PL measurements using a streak camera with an excitation pulse centered at 633 nm (Methods). Figure 1e is the PL intensity plot as a function of energy and time. By integrating the detected intensity over wavelength within the region of interest outlined by the grey vertical lines in Fig. 1e, we obtain the time trace of the PL presented in Fig. 1f. From fitting the PL time trace with a monoexponential function, we find the exciton lifetime to be 11 ps. This corresponds to an intrinsic linewidth of 30 µeV, smaller than the measured linewidth of about 350 µeV. This comparison shows that the narrow exciton PL linewidth is dominated by pure dephasing or inhomogeneous broadening.

The observed exciton photoluminescence is highly anisotropic. As shown in Fig. 1d, the photoluminescence is dominated by emission along a single direction. The degree of linear polarization is represented as $\rho = \frac{I_\perp - I_\parallel}{I_\perp + I_\parallel}$, where $I_\parallel$ ($I_\perp$) is the peak intensity of horizontally (vertically) polarized photoluminescence. The value of $\rho$ approaches unity (0.96). This observation of a highly linearly polarized exciton provides the first indication that the exciton couples strongly to the zigzag antiferromagnetic order. Even though the three-fold rotational symmetry is already broken by the monoclinic layer stacking, the observed optical anisotropy is of a purely magnetic origin, as we will discuss below. We have also performed polarization-dependent excitation and unpolarized detection measurements. These measurements show that photoluminescence has little dependence on the excitation polarization (Extended Data Fig. 2).

The strong exciton-magnetic order coupling is corroborated by the temperature dependent photoluminescence (Extended Data Fig. 3). As temperature increases, the photoluminescence intensity drops while the peak width broadens (Fig. 1d). Above 100 K, the exciton emission loses its Lorentzian lineshape and becomes increasingly broad as the temperature approaches the Néel temperature, nearly vanishing above the Néel temperature of 150 K[20]. Although thermal disassociation and phonon effects are the primary factors that influence the temperature dependent photoluminescence linewidth and intensity, enhanced spin fluctuations and changes in the electronic structure near the Néel temperature could be contributing factors as well. There is also an overall concomitant red shift of the exciton peak as temperature increases, which is consistent with the reduced bandgap that results from lattice warming effects. Notably, the photoluminescence remains strongly linearly polarized up to about 70 K, but the degree of polarization decreases and eventually vanishes above the Néel temperature (Inset Fig. 1d). The exciton linear polarization dependence on temperature resembles the in-plane magnetic susceptibility anisotropy, pointing to an intimate connection between them (Extended Data Figure 4). This connection will be further strengthened by the linear dichroism measurement presented later.

Figure 2a shows the photoluminescence as a function of layer number at 15 K. The exciton luminescence can be clearly observed down to the bilayer, but there is no appreciable photoluminescence from monolayers. There is a strong reduction of the photoluminescence intensity commensurate with reduced layer number. Figure 2b shows the photoluminescence mapping of the sample area highlighted in Fig. 1b, where the 7 and 4 layer regions are bright while the adjacent thinner layers are barely visible. Furthermore, the photoluminescence peak blue shifts relative to the bulk photoluminescence peak as the layer number decreases (Fig. 2c). All these effects manifest most strongly in the transition from trilayer to bilayer.

The observation of thickness dependent photoluminescence properties implies layer-dependent electronic structures. We have performed a first-principles calculation from monolayer to six layers (Extended Data Figure 5). The calculation shows that the valence band maximum (VBM) lies between the Γ and X points of the Brillouin zone. In trilayer and above, the conduction band minimum (CBM) also lies in this region, though it rests closer to the Γ point, making the interband transition indirect with a relatively small CBM-VBM crystal momentum separation. By contrast, in bilayer and monolayer, the CBM moves to the Y point, drastically increasing the momentum separation of the indirect bandgap. Furthermore, the calculation shows an increased bandgap as the layer number decreases. Both of these theoretical features are corroborated by our data. The increase in crystal momentum separation of the interband transition explains the reduction in PL intensity with decreasing thickness, particularly going from 3L to 2L, and the increased bandgap in thinner flakes explains the observed blue-shift of the PL peaks. Our calculation is consistent with the very recent report on layer-dependent electronic structure in $NiPS_3$[5]. The observed, sensitive dependence of excitonic photoluminescence properties on the evolution of band structure versus layer number points to the Wannier-like character of the exciton wavefunction.

The photoluminescence linewidth broadens slightly from five layer to trilayer but increases by a factor of 10 from trilayer to bilayer. This sudden linewidth increase cannot be explained solely by the possible substrate sensitivity at the atomic thin limit. In fact, samples with and without hBN encapsulation do not show any appreciable differences in the linewidth (Extended Data Fig. 6). We speculate that the enhanced spin fluctuations at the atomically thin limit[26] may also contribute to the linewidth broadening in bilayer.

The exciton photoluminescence in atomically thin flakes is also highly linearly polarized. Here, we use a five-layer sample to illustrate this property due to its strong photoluminescence. Figure 3a shows the photoluminescence intensity plot as function of linear polarization detection angle (see Extended data Fig. 6 for thinner samples). Similar to what was discussed in the thin bulk crystals, the exciton photoluminescence is strongly polarized along the vertical axis with a near unity $\rho$, again pointing towards strong coupling of the exciton to the zigzag antiferromagnetic order (Fig. 3b). Figure 3c plots the temperature-dependent exciton photoluminescence with vertically polarized detection (see Extended Data Fig. 7 for the temperature dependence of the extracted linewidth and intensity). Evidently, the peak intensity first reduces slowly as temperature increases, but then quickly drops as the temperature rises above 40 K and is barely visible above 100 K. The degree of linear polarization $\rho$ has similar temperature dependence (Fig. 3c inset). The exact cause of the disappearance of the photoluminescence signal in the 5L at a temperature lower than that in thin bulk crystal is not clear. The data show that, for atomically thin flakes, linearly polarized PL measurements can be used to study the zigzag AFM and the associated optical anisotropy, but they have a limitation in probing the temperature dependent magnetic properties, especially near the Néel temperature.

In addition to photoluminescence measurements, we perform linear polarization resolved optical reflection measurements. We define linear dichroism (LD) as $\frac{R_\perp - R_\parallel}{R_\perp + R_\parallel}$, where $R_\parallel$ ($R_\perp$) is the peak intensity of horizontally (vertically) polarized optical reflection (see methods). Figure 4a shows a broad LD spectrum from a thin bulk sample. The LD is clearly enhanced at charge transfer exciton resonances (e.g. at 2.2 eV[27]). In addition to these broad peaks, there is a sharp resonance at 838 nm (or 1.480 eV), about 5 meV above the anisotropic exciton observed in the photoluminescence. The line shape of the peak results from the combination of thin film interference and a nearby state with weaker anisotropy (Extended Data Fig. 8).

LD strongly depends on the temperature. As shown in the inset of Fig. 4a, LD first decreases slowly as the temperature increases, then drops drastically as the temperature approaches the Néel temperature and vanishes above it. This temperature dependence is similar to the previously discussed linear polarization of the exciton photoluminescence. In addition, the LD axis is along the same optical anisotropy axis as photoluminescence and Raman scattering (Extended Data Figure 1). These results further support the fact that both LD and linearly polarized exciton photoluminescence originate from the zigzag antiferromagnetic order, which breaks three-fold rotation symmetry (Extended Data 5). Both LD and photoluminescence also have little dependence on out of plane magnetic field up to 9T, consistent with the strong antiferromagnetic exchange interactions[24,25] (Extended Data Fig. 9). Note that the degree of linear polarization in exciton photoluminescence is much larger than LD. This discrepancy is due to the fact that LD is measured on top of a strong unpolarized reflection background, while exciton photoluminescence is nearly background-free.

On the high-energy side of the sharp exciton peak, there are over 10 periodic oscillation features (Fig. 4b). Fourier transform reveals a period of about 28 meV. The oscillation features vanish on bare $SiO_2$/Si substrate, which rules out artifacts from the experimental setup as the source of the oscillations. Moreover, the period of the oscillations is independent of the layer thickness, which precludes simple interference effects as the origin. Instead, the high-energy periodic spectral structures are likely exciton-phonon bound states[31,32]. Such states have previously been observed in molecular crystals[33]. The energy separation between the exciton and the bound states is usually 10%-20% smaller than the corresponding phonon energy[31-35]. By examining the Raman spectrum (Extended Data Fig. 1), we find the $A_{1g}$ phonon at 32 meV (258 cm$^{-1}$) to be the most likely candidate bound to the exciton.

Compared to exciton-phonon bound states observed in various bulk crystals, there are several unique features in our observation. First, usually only one or up to a few exciton-phonon bound states are observed in reported systems[31-33]. Remarkably, we observe over 10 states, signifying an exceptionally strong exciton-phonon coupling. Furthermore, $NiPS_3$ is a charge-transfer insulator where the nickel 3$d$ orbitals and sulfur $p$ orbitals form a ligand field charge transfer transition. The involved $A_{1g}$ phonon represents the out-of-plane motion of the S atoms (Fig. 4f). Such a motion would modify the S-Ni bond. Therefore, the data suggest that the $A_{1g}$ mode strongly modifies the charge-transfer energy, and thus the electronic structure of the crystal. This observation, as well as the strong thickness dependence of the PL intensity, implies that the observed exciton is not a Frenkel exciton localized in the $Ni^{2+}$ cation[36,37].

Secondly, the exciton-phonon bound states can be observed down to the atomically thin limit. Figure 4c shows the LD spectra versus the layer number. Excitons with associated $A_{1g}$ phonon bound states can be clearly resolved down to trilayers. However, the signal becomes too weak to resolve in both bilayers and monolayers. Note that the oscillation amplitude appears to be larger for higher-order bound states. This is consistent with the fact that higher-order bound states involve more phonons and thus larger anisotropy, yielding larger LD signal than the lower order ones.

Lastly, the exciton-phonon bound states strongly couple to the magnetic order. Figure 4d shows the LD intensity plot as a function of temperature and photon energy. The exciton-phonon bound states are evident below the Néel temperature. The LD amplitude of the bound states decreases as the temperature increases. This observation is highlighted by the LD spectra at selected temperatures (Fig. 4e). The corresponding Fourier transform of the data is plotted in the inset. The oscillation amplitude vanishes above the Néel temperature, consistent with the exciton's strong coupling to the zigzag antiferromagnetic order. The oscillation frequency shows a slight blue shift as temperature increases, signifying the weakening of the exciton-phonon bonding effect.

In summary, our work reveals $NiPS_3$ as a new addition to the family of 2D semiconductors that hosts robust excitons down to the bilayer. $NiPS_3$ excitons are highly anisotropic with the optical anisotropy axis determined by the zigzag antiferromagnetic order. Due to the vanishing net magnetization of antiferromagnetic order, nanoscale antiferromagnetism is usually challenging to probe. The demonstrated simple optical anisotropy measurements thus provide a new and reliable optical means to study a broad class of rotational-symmetry-breaking antiferromagnetic orders in correlated materials, such as zigzag and stripe antiferromagnetism. These excitons also exhibit strong dependence on the layer thickness, which originates from layer-dependent electronic structure. This layer-thickness-dependent property is distinct from the photoluminescence from highly localized molecular orbital transitions, such as the photoluminescence of the *d-d* transition observed in atomically thin $CrI_3$[4]. Rather, the exciton in $NiPS_3$ behaves more like a Wannier exciton. These properties make NiPS$_3$ amenable to band structure engineering via stacking into twisted homobilayers and moiré heterostructures with different van der Waals materials[39,40]. We thus foresee that NiPS$_3$ can be useful for engineering magnetic moiré excitons and exploring excitonic many-body quantum states. Lastly, exceptionally strong exciton-phonon coupling points to the possibility of strong exciton magnon coupling as well as tuning magnetic order and correlated effects in NiPS$_3$ via optical manipulation of lattice degree of freedom.

**Methods:**

**Crystal growth and sample fabrication**: Single crystals of NiPS$_3$ were synthesized by chemical vapor transport (CVT) method using iodine as the transport agent. Stoichiometric amounts of nickel powder (99.998%), phosphorous powder (98.9%) and sulfur pieces (99.9995%) were mixed with iodine (1 mg/cc) and sealed in quartz tubes (10 cm in length) under high vacuum. The tubes were placed in a horizontal one-zone tube furnace with the charge near the center of the furnace. Sizeable crystals were obtained after gradually heating the precursor up to 950°C, dwelling for a week, and cooling down to room temperature. The crystals were then exfoliated onto (90nm) SiO2/Si substrate in an Argon gas protected glove box. The layer thickness was identified by atomic force microscopy imaging in the glove box as well as by optical contrast analysis. To protect the samples from degrading in ambient conditions, the samples were sealed in a copper sample holder under a transparent cover slip inside the glovebox before being transported and loaded into the cryostat. The total time the sealed sample spent outside of the glove box prior to the loading was usually less than 10 minutes.

**Photoluminescence and Raman measurement**: A HeNe laser (633 nm) was used to excite the NiPS$_3$ sample, which was placed in a closed cycle cryostat with temperature range from 5K to 300K. Both photoluminescence and Raman measurements were polarization-resolved and collected by a spectrometer with a liquid nitrogen cooled CCD camera. A magnetic field up to 9T was applied in Faraday geometry.

**Linear dichroism (LD) spectroscopy**: The measurements were carried out in the reflection geometry. A supercontinuum laser with a filter set was utilized as a tunable light source with 1 nm spectral resolution. The laser beam was double modulated by a photoelastic modulator (PEM) with a maximum retardance of $\lambda/2$ and a mechanical chopper. After phase modulation, the light passed through a half-waveplate, and then was focused down onto the sample at normal incidence with an objective lens. A laser power of 5 µW was used. The reflected light was detected by a photodiode, and further demodulated at 100 kHz and 1 kHz, which corresponds to the PEM linear polarization modulation and the chopper modulation frequency, respectively.

**Band structure calculation**: The Quantum Espresso package[38] was used to perform to the first-principles calculations by taking the Rappe-Rabe-Kaxiras-Joannopoulos ultrasoft pseudoptentials for the semi-local Perdew-Burke-Ernzerhof (PBE) generalized gradient approximation (GGA) and the VdW-D2 correction. The crystal structures were relaxed with the convergence threshold $10^{-4}$ Ry for total energy and $10^{-4}$ Ry/Bohr for force. The 6×3×1 Monkhorst-Pack grids were used to sample the k points during the self-consistency calculation for the ground state electron structure, which is converged below the criterion $10^{-7}$ Ry for total energy. The electron correlation effect was taken into account by the GGA+U method by setting U=6.45 eV[21].

**Time-resolved PL measurements:** Time-resolved PL measurements were conducted using a streak camera (Hamamatsu-C10910) synchronized with an oscillator (Coherent-Mira 900-F). Laser pulses from the oscillator with a central wavelength of 820 nm and temporal width of ~150 fs were used to pump a nonlinear photonic crystal fiber (NKT Photonics-FemtoWhite 800) to generate supercontinuum pulses. Several bandpass filters were then used to center the excitation bandwidth around 633 nm. The excitation was focused by an objective lens onto the sample, which was held at 5 K by a closed cycle cryostat (Attocube-Attodry 2100). The PL signal was spectrally and temporally dispersed within the streak camera and detected on the internal CCD. Since the measured IRF was significantly shorter than the PL time trace, the exciton lifetime was obtained by fitting the time trace with a monoexponential function.

**References:**

1. S. Schmitt-Rink *et al.* "Linear and nonlinear optical properties of semiconductor quantum wells", *Advances in Physics* **38**, 89(1989).
2. K. F. Mak & J. Shan. "Photonics and optoelectronics of 2D semiconductor transition metal dichalcogenides", *Nature Photonics* **10**, 216(2016).
3. K. S. Burch *et al.* "Magnetism in two-dimensional van der Waals materials", *Nature* **563**, 47(2018).
4. K. L. Seyler *et al.* "Ligand-field helical luminescence in a 2D ferromagnetic insulator", *Nature Physics* **14**, 277(2018).
5. Christopher Lane & J.-X. Zhu. "Thickness dependence of electronic structure and optical properties of a correlated van der Waals antiferromagnet NiPS3 thin film", *arXiv:2003.01614* (2020).
6. J. K. Furdyna. "Diluted magnetic semiconductors", *Journal of Applied Physics* **64**, R29(1988).
7. H. Ohno *et al.* "Magnetotransport properties of p-type (In,Mn)As diluted magnetic III-V semiconductors", *Physical Review Letters* **68**, 2664(1992).
8. H. Ohno *et al.* "(Ga,Mn)As: A new diluted magnetic semiconductor based on GaAs", *Applied Physics Letters* **69**, 363(1996).
9. K. S. Burch *et al.* "Optical properties of III-Mn-V ferromagnetic semiconductors", *Journal of Magnetism and Magnetic Materials* **320**, 3207(2008).
10. T. Dietl & H. Ohno. "Dilute ferromagnetic semiconductors: Physics and spintronic structures", *Reviews of Modern Physics* **86**, 187(2014).
11. H. Togashi *et al.* "Field-Induced Davydov Splitting of Excitons in Quasi-One-Dimensional Antiferromagnet CsMnCl3·2H2O", *Journal of the Physical Society of Japan* **57**, 353(1988).
12. J. P. van der Ziel. "Davydov Splitting of the $^2$E Lines in Antiferromagnetic $Cr_2O_3$", *Physical Review Letters* **18**, 237(1967).
13. W. Heiss *et al.* "Giant tunability of exciton photoluminescence emission in antiferromagnetic EuTe", *Physical Review B* **63**, 165323(2001).
14. S. L. Gnatchenko *et al.* "Exciton-magnon structure of the optical absorption spectrum of antiferromagnetic MnPS3", *Low Temperature Physics* **37**, 144(2011).
15. J. A. Gaj, & J. Kossut. "Introduction to the Physics of Diluted Magnetic Semiconductors; " *Springer Series in Materials Science 144; Springer–Verlag: Berlin*(2010).

| 16 | B. Huang *et al.* "Layer-dependent ferromagnetism in a van der Waals crystal down to the monolayer limit", *Nature* **546**, 270(2017). |
|---|---|
| 17 | C. Gong *et al.* "Discovery of intrinsic ferromagnetism in two-dimensional van der Waals crystals", *Nature* **546**, 265(2017). |
| 18 | Z. Zhang *et al.* "Direct Photoluminescence Probing of Ferromagnetism in Monolayer Two-Dimensional CrBr3", *Nano Letters* **19**, 3138(2019). |
| 19 | M. Wu *et al.* "Physical origin of giant excitonic and magneto-optical responses in two-dimensional ferromagnetic insulators", *Nature Communications* **10**, 2371(2019). |
| 20 | G. Le Flem *et al.* "Magnetic interactions in the layer compounds MPX3 (M = Mn, Fe, Ni; X = S, Se)", *Journal of Physics and Chemistry of Solids* **43**, 455(1982). |
| 21 | B. L. Chittari *et al.* "Electronic and magnetic properties of single-layer $MPX_3$ metal phosphorous trichalcogenides", *Physical Review B* **94**, 184428(2016). |
| 22 | J.-U. Lee *et al.* "Ising-Type Magnetic Ordering in Atomically Thin FePS3", *Nano Letters* **16**, 7433(2016). |
| 23 | X. Wang *et al.* "Raman spectroscopy of atomically thin two-dimensional magnetic iron phosphorus trisulfide ($FePS_3$) crystals", *2D Materials* **3**, 031009(2016). |
| 24 | A. R. Wildes *et al.* "Magnetic structure of the quasi-two-dimensional antiferromagnet $NiPS_3$", *Physical Review B* **92**, 224408(2015). |
| 25 | D. Lançon *et al.* "Magnetic exchange parameters and anisotropy of the quasi-two-dimensional antiferromagnet $NiPS_3$", *Physical Review B* **98**, 134414(2018). |
| 26 | K. Kim *et al.* "Suppression of magnetic ordering in XXZ-type antiferromagnetic monolayer NiPS3", *Nature Communications* **10**, 345(2019). |
| 27 | S. Y. Kim *et al.* "Charge-Spin Correlation in van der Waals Antiferromagnet $NiPS_3$", *Physical Review Letters* **120**, 136402(2018). |
| 28 | C.-T. Kuo *et al.* "Exfoliation and Raman Spectroscopic Fingerprint of Few-Layer NiPS3 Van der Waals Crystals", *Scientific Reports* **6**, 20904(2016). |
| 29 | F. Cadiz *et al.* "Excitonic Linewidth Approaching the Homogeneous Limit in MoS2 Based van der Waals Heterostructures", *Physical Review X* **7**, 021026(2017). |
| 30 | J. Wierzbowski *et al.* "Direct exciton emission from atomically thin transition metal dichalcogenide heterostructures near the lifetime limit", *Scientific Reports* **7**, 12383(2017). |
| 31 | Y. Toyozawa & J. Hermanson. "Exciton-Phonon Bound State: A New Quasiparticle", *Physical Review Letters* **21**, 1637(1968). |
| 32 | R. Merlin *et al.* "Multiphonon processes in YbS", *Physical Review B* **17**, 4951(1978). |
| 33 | Y. B. Levinson & E. I. Rashba. "Electron-phonon and exciton-phonon bound states", *Reports on Progress in Physics* **36**, 1499(1973). |
| 34 | W. Y. Liang & A. D. Yoffe. "Transmission Spectra of ZnO Single Crystals", *Physical Review Letters* **20**, 59(1968). |
| 35 | W. C. Walker *et al.* "Phonon-Induced Splitting of Exciton Lines in MgO and BeO", *Physical Review Letters* **20**, 847(1968). |
| 36 | E. J. K. B. Banda. "Optical absorption of NiPS3 in the near-infrared, visible and near-ultraviolet regions", *Journal of Physics C: Solid State Physics* **19**, 7329(1986). |
| 37 | M. Piacentini *et al.* "Optical transitions, XPS, electronic states in NiPS3", *Chemical Physics* **65**, 289(1982). |

38   P. Giannozzi *et al.* "QUANTUM ESPRESSO: a modular and open-source software project for quantum simulations of materials", *Journal of Physics: Condensed Matter* **21**, 395502(2009).
39   K. Hejazi *et al.* "Noncollinear phases in moiré magnets", *Proceedings of the National Academy of Sciences* **117,** 10721 (2020).
40   M. Onga *et al.* "Antiferromagnet–Semiconductor Van Der Waals Heterostructures: Interlayer Interplay of Exciton with Magnetic Ordering", *Nano Letters* **20**, 4625 (2020).

**Acknowledgements:** This work was mainly supported by the Department of Energy, Basic Energy Sciences, Materials Sciences and Engineering Division (DE-SC0012509 and DE-SC0018171). Device fabrication and part of photoluminescence measurement were supported by Air Force Office of Scientific Research (AFOSR) Multidisciplinary University Research Initiative (MURI) program, grant no. FA9550-19-1-0390. Bulk crystal growth is supported by NSF MRSEC DMR-1719797 and the Gordon and Betty Moore Foundation's EPiQS Initiative, Grant GBMF6759 to JHC. YW is supported by NSFC Projects (Grant Nos. 61674083, 11604162). The authors also acknowledge the use of the facilities and instrumentation supported by NSF MRSEC DMR-1719797. XX and JHC acknowledges the support from the State of Washington funded Clean Energy Institute.

**Author contributions:** XX, QZ and KH conceived the experiment. KH and QZ fabricated samples and performed optical measurements, assisted by JF, and GD. All authors contributed to the data analysis and interpretation. YW and WY performed band structure calculation. CW and DX calculated the optical anisotropy. QJ and JHC synthesized and characterized the bulk crystals. KH, QZ, XX, and JF wrote the paper with input from all authors. All authors discussed the results.

**Competing Interests:** The authors declare no competing financial interests.

**Data Availability:** The datasets generated during and/or analysed during this study are available from the corresponding author upon reasonable request.

# Figures

**Figure 1 | Anisotropic antiferromagnetic excitons in thin bulk NiPS$_3$ crystal. a**, Schematic of zigzag antiferromagnetic order in a single layer of NiPS$_3$ and of monoclinic stacking with ferromagnetic interlayer coupling in bulk crystal. **b**, Optical microscope image of exfoliated samples. **c**, Atomic force microscopy image of the area enclosed by the dashed rectangle in (b). Layer numbers are indicated. **d**, Linear polarization resolved photoluminescence of a thin bulk NiPS$_3$ at selected temperatures. Blue and red data represent horizontally (H) and vertically (V) polarized detection. Inset: degree of linear polarization as a function of temperature. Error bars were obtained from a Lorentzian fit of the PL signals. **e,** Raw streak camera image of time resovled photoluminescence at 5 K. **f,** Integrated intensity time trace (red curve) of the area enclosed by the dashed lines in (e). Blue curve is the measured instrument response function. The exciton lifetime of ~11 ps was extracted by fitting the time trace with a monoexponential function.

**Figure 2 | Thickness-dependent exciton luminescence and electronic structure. a,** Exciton photoluminescence as a function of layer number. Both excitation and detection are vertically polarized. **b**, Photoluminescence intensity map of the area enclosed by the white dashed box in Fig. 1b. **c,** Extracted peak energy position vs layer number. Error bars were obtained from a Lorentzian fit of the PL signals.

**Figure 3 | Exciton-zigzag antiferromagnetic order coupling in a five-layer sample. a**, Photoluminescence intensity plot as a function of linear polarization detection angle. **b**, Horizontally (H) and vertically (V) polarized photoluminescence spectra. **c**, temperature dependent photoluminescence with co-vertically polarized excitation and detection. Inset: degree of linear polarization vs temperature. Error bars were obtained from a Lorentzian fit of the PL signals.

**Figure 4 | Multiple exciton-phonon bound states. a**, Linear dichroism (LD) spectrum of a thin bulk flake. Data is taken at 5 K. Inset: Temperature-dependent LD extracted at photon energy of 2.17 eV. Error bars were obtained from a sinusoidal fit of the rotational LD measurements. **b**, Low energy Zoom in LD spectrum. **c**, LD vs layer number at atomically thin limit. Data is taken at 5 K. Spectra are shifted for clarity. **d**, LD intensity plot as a function of temperature and photon energy. **e**, LD spectra at selected temperatures. Inset: Corresponding Fourier transformation of LD spectra. **f**, Cartoon of exciton-phonon coupling via A$_{1g}$-mode modulated ligand field transition. Direction of vibrational motion of S atoms depicted with in-to-figure (×) and out-of-figure (•) symbols.

# Figure 1

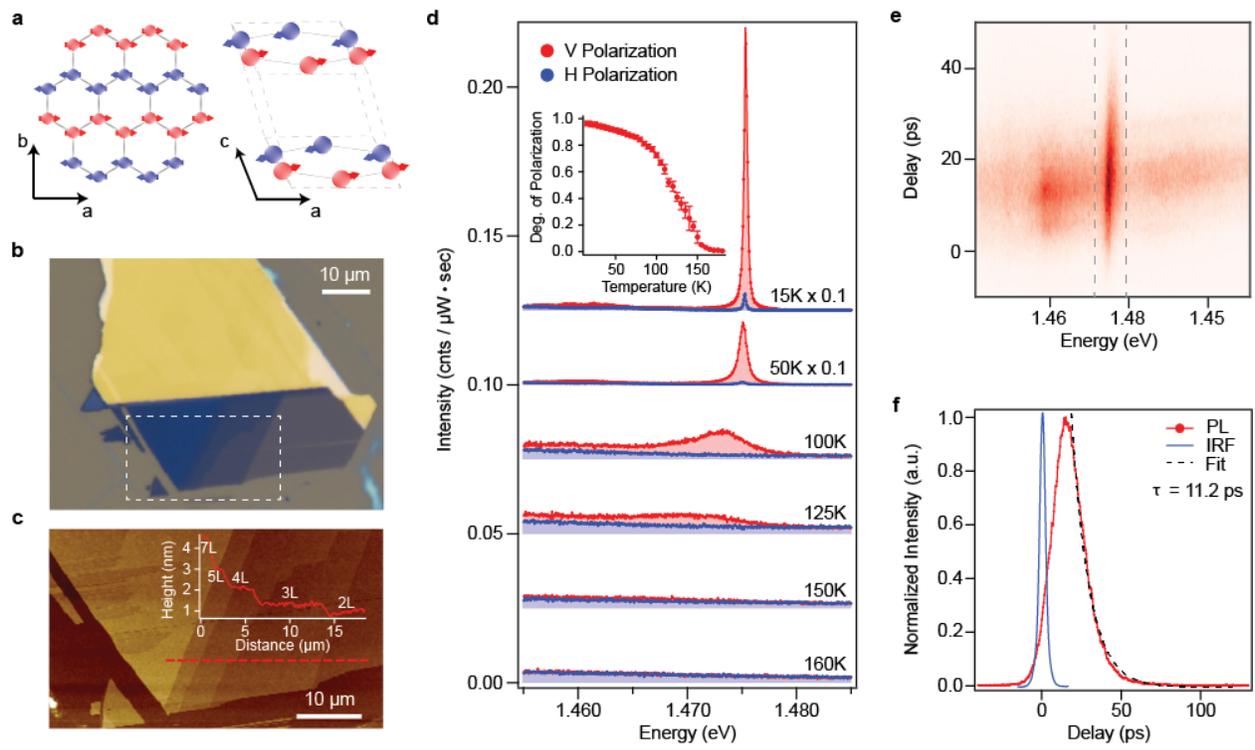

**Figure 2**

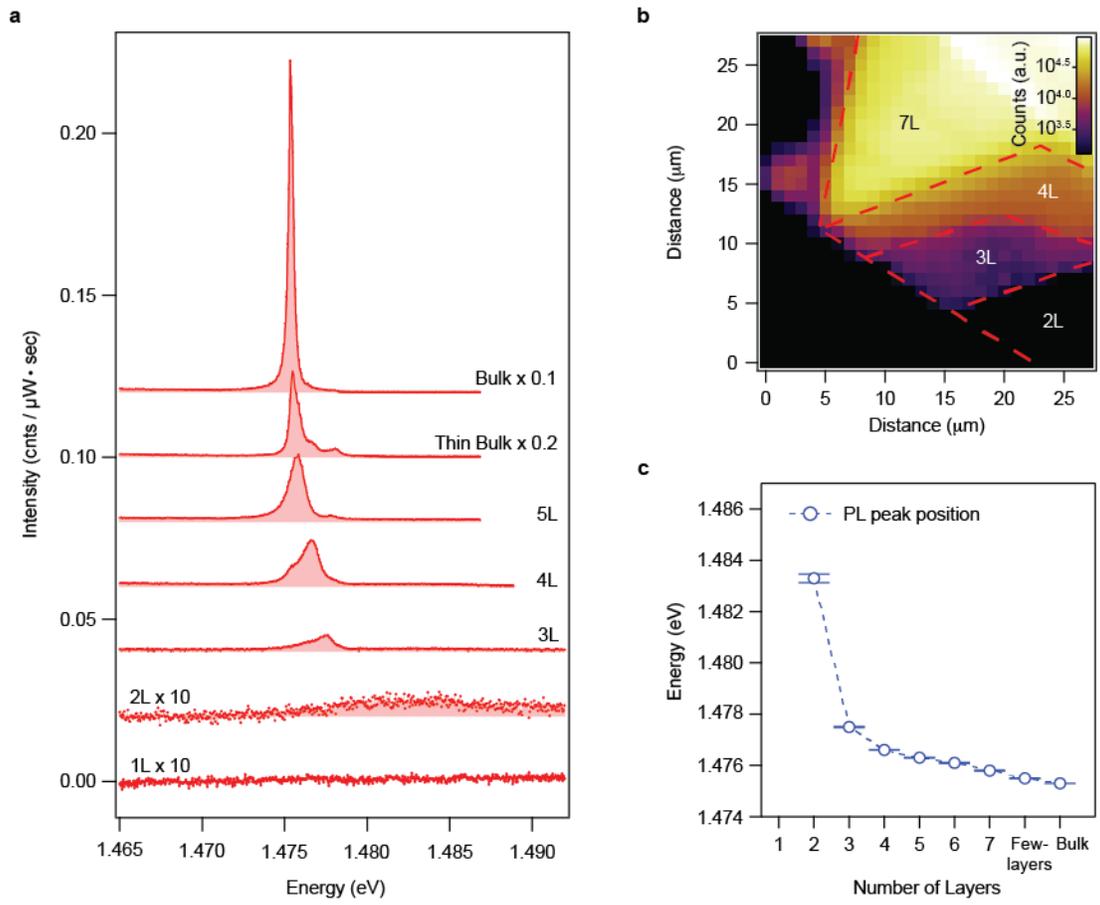

**Figure 3**

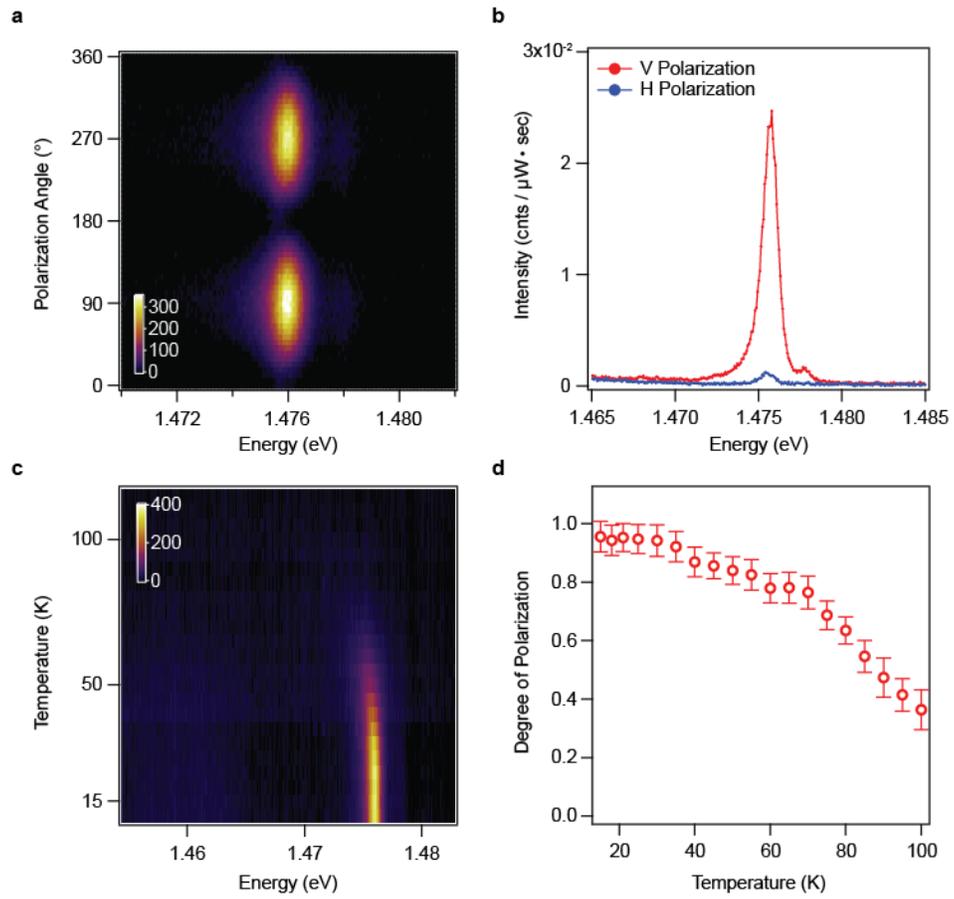

**Figure 4**

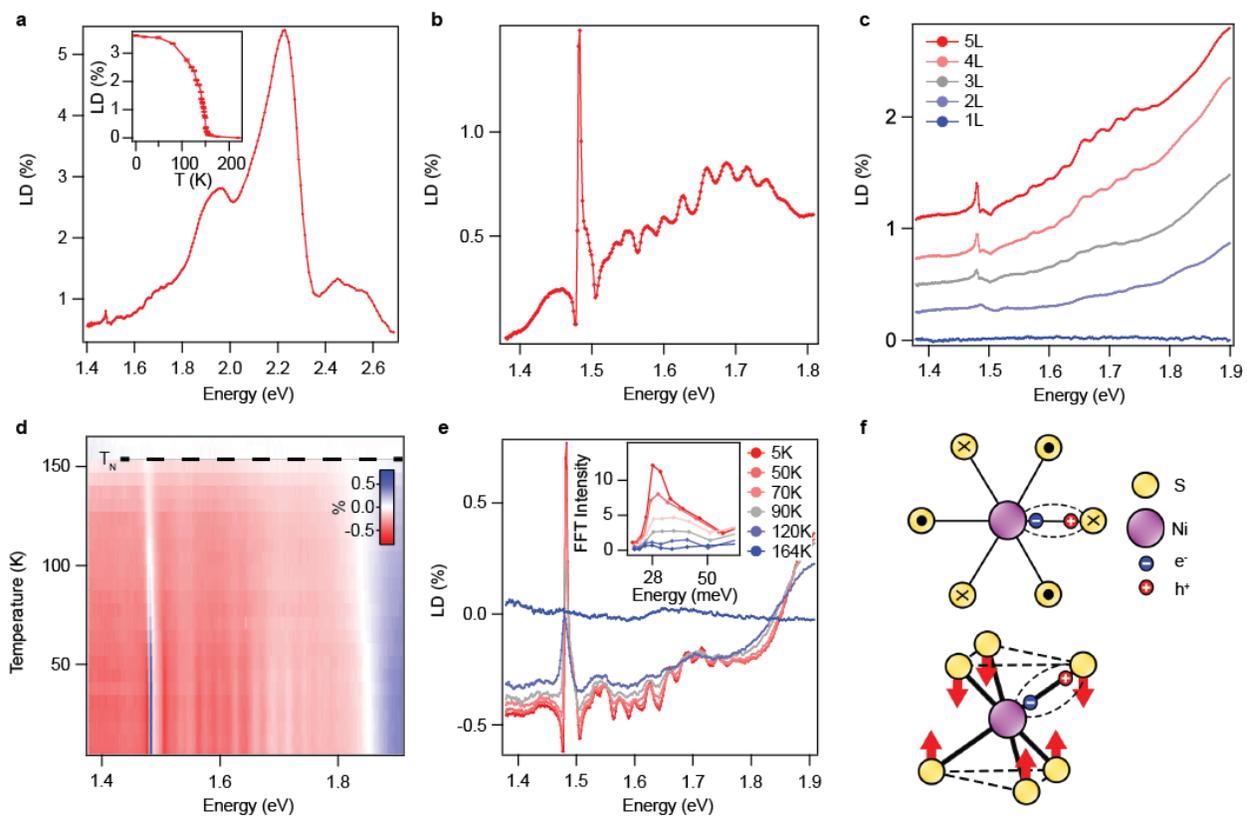

**Extended Data**

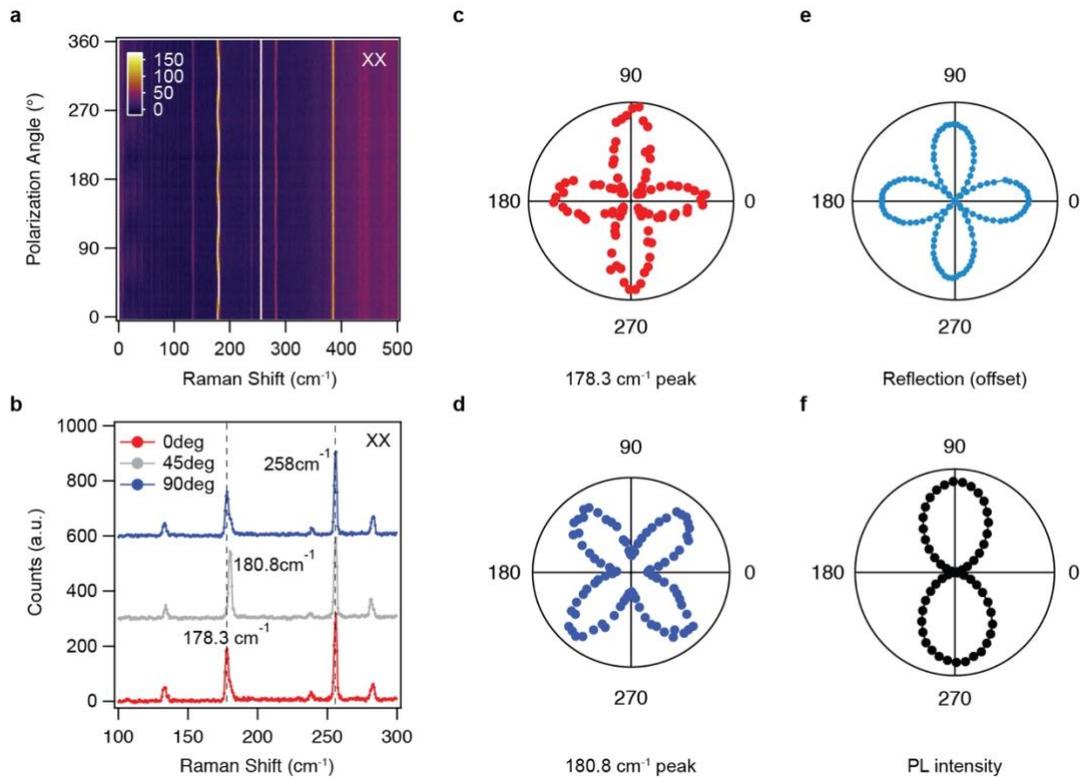

**Extended Data Figure 1 | Polarization-resolved Raman spectroscopy, optical reflection and photoluminescence measurements of the same thin-bulk NiPS$_3$ flake. a**, Co-linearly polarized Raman scattering (the XX channel). **b**, Raman spectra at 0º, 45º, and 90º of linear polarization. **c**, Polar plot of 178.3 cm$^{-1}$, and **d**, 180.8 cm$^{-1}$ Raman mode intensity as function of linear polarization angle. Data is taken at 15 K. **e**, Offset optical reflection at 633nm as function of linear polarization angle. The positive lobes indicate the high reflection direction. **f**, Polar plot of photoluminescence intensity as a function of linear polarization detection angle. Near unity linearly polarized photoluminescence is observed along the vertical direction. Horizontal (H) and vertical (V) polarization direction is defined along 0º and 90º in this figure, respectively.

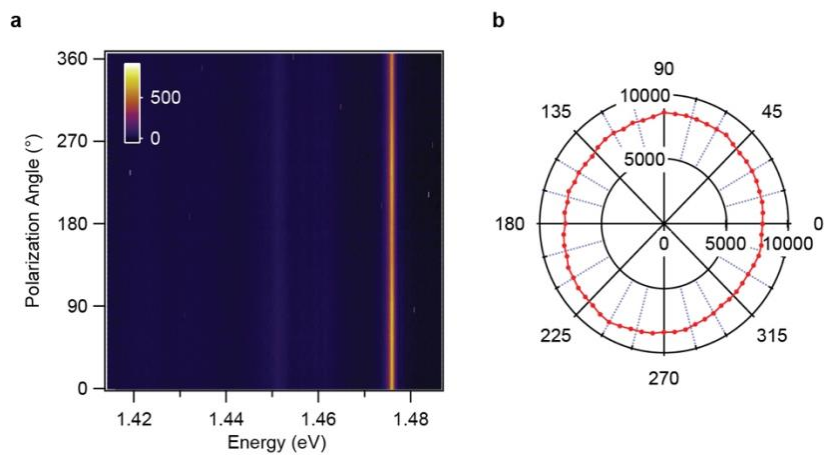

**Extended Data Figure 2 | Photoluminescence as a function of excitation linear polarization angle. a,** Photoluminescence intensity plot as a function of photon energy and excitation polarization. Detection is unpolarized. **b,** Polar plot of peak intensity vs excitation polarization.

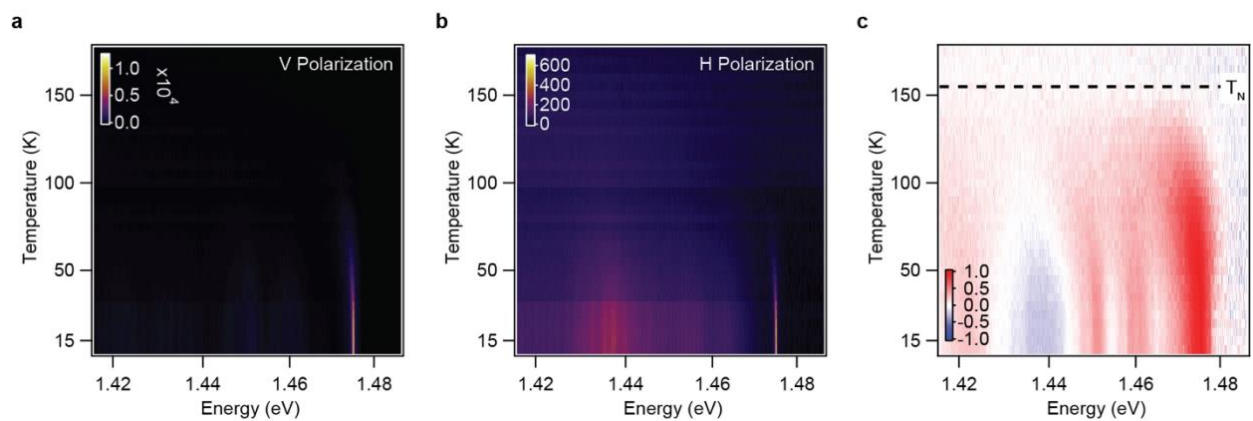

**Extended Data Figure 3 | Temperature- and polarization-resolved photoluminescence from a bulk NiPS$_3$ flake.** The excitation laser is at 633 nm and vertically polarized. Photoluminescence intensity plot with **a,** vertically and **b,** horizontally polarized detection. The main excitonic peak is strongly linearly polarized. There are low energy and very weak photoluminescence features. These states are more visible in **c,** temperature dependent degree of linear polarization plot. The first three features have energy separation of about 10~12 meV, implying that the two low energy features are phonon replica of the exciton. The feature at 1.44 eV has stronger photoluminescence intensity than the first two while having opposite polarization, indicating that it has a different origin from the first two. We speculate that this state may arise from the electric dipole forbidden *d-d* transitions localized in the Ni$^{2+}$. The identification of the exact origin of these states requires future efforts.

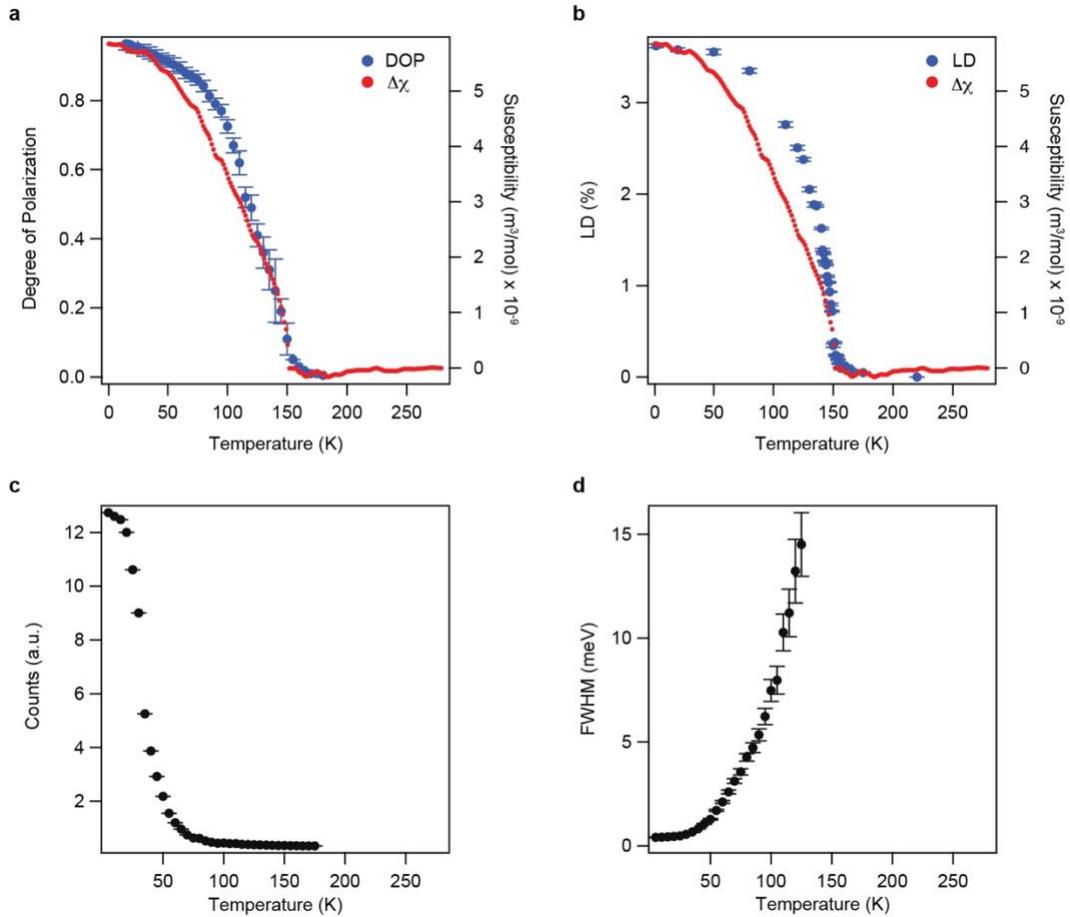

**Extended Data Figure 4 | Comparison of temperature dependent exciton photoluminescence properties with in-plane magnetic susceptibility anisotropy Δχ.** Δχ is the magnetic susceptibility difference between *a* and *b* axes of bulk crystal NiPS$_3$, extracted from Ref. 24. The photoluminescence measurements are done on the sample presented in Fig. 1e in the main text. **a** and **b** overlays Δχ (red) with degree of linear polarization (DOP, blue) and linear dichroism (LD, blue) as a function of temperature, respectively. The temperature dependent behavior of DOP and LD resembles that of Δχ, supporting the in-plane magnetic susceptibility anisotropy as their cause. Error bars for DOP were obtained from a Lorentzian fit of the PL signals. Temperature dependent **c,** photoluminescence intensity and **d,** full width at half maximum (FWHM). FWHM is not shown above 120 K as the peak shape was not suitable for fitting. Error bars in **c** and **d** signify the confidence bounds of a Lorentzian fit.

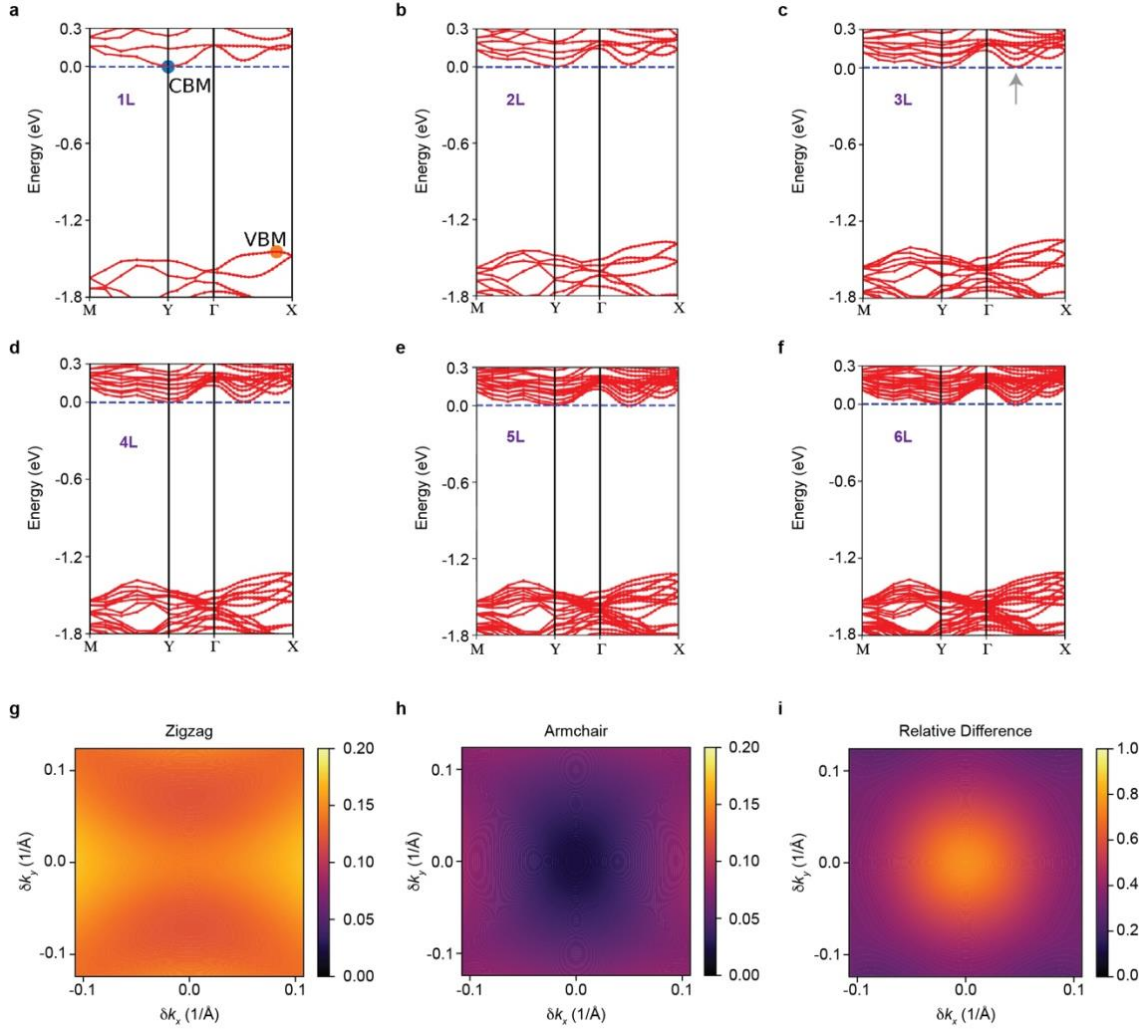

**Extended Data Figure 5 | DFT calculated band structure and oscillator strength of inter-band optical transition. a-f,** layer number dependent band structure. The rectangular unit cell and Brillouin zone have been exploited during the calculations. The k path for the band is along M-Y-Γ-X, where Γ=(0,0), X=(π/a,0),Y=(0,π/b), M=(π/a,π/b). The zero energy has been set to the conduction band minimum at Y point for comparison. The highest energy valence band is near the X point, indicated in the figure. We can see that while the lowest energy conduction band is at Y point in monolayer and bilayer, an energy minimum between Γ and X, indicated by the arrow in the trilayer band structure, appears as the layer number increases to 3 and above. Therefore, $NiPS_3$ experiences a transition from indirect in monolayer and bilayer to less indirect bandgap in trilayer and above. This transition is likely responsible for the observed layer thickness dependent exciton photoluminescence, which manifest most strongly between trilayer and bilayer. **g-i,** Oscillator strength along the zigzag (*a*) and armchair (*b*) direction in the vicinity of the X point. We have averaged the oscillator strength over the quasi-degenerate bands near the band edges, i.e., $\frac{1}{32}\sum_{v,c}|\langle ck|\hat{v}|vk\rangle|^2$, where *v* runs over the 4 near degenerate valence bands and *c* runs over 8 nearly degenerate conduction bands. The unit of the averaged velocity matrix elements is $(eV \cdot Å)^2$. The large difference of the oscillator strength between zigzag and armchair, i.e. (zigzag-armchair)/(zigzag+armchair), represents highly anisotropic states due to the zigzag antiferromagnetic order. This is consistent with the observed exciton LD.

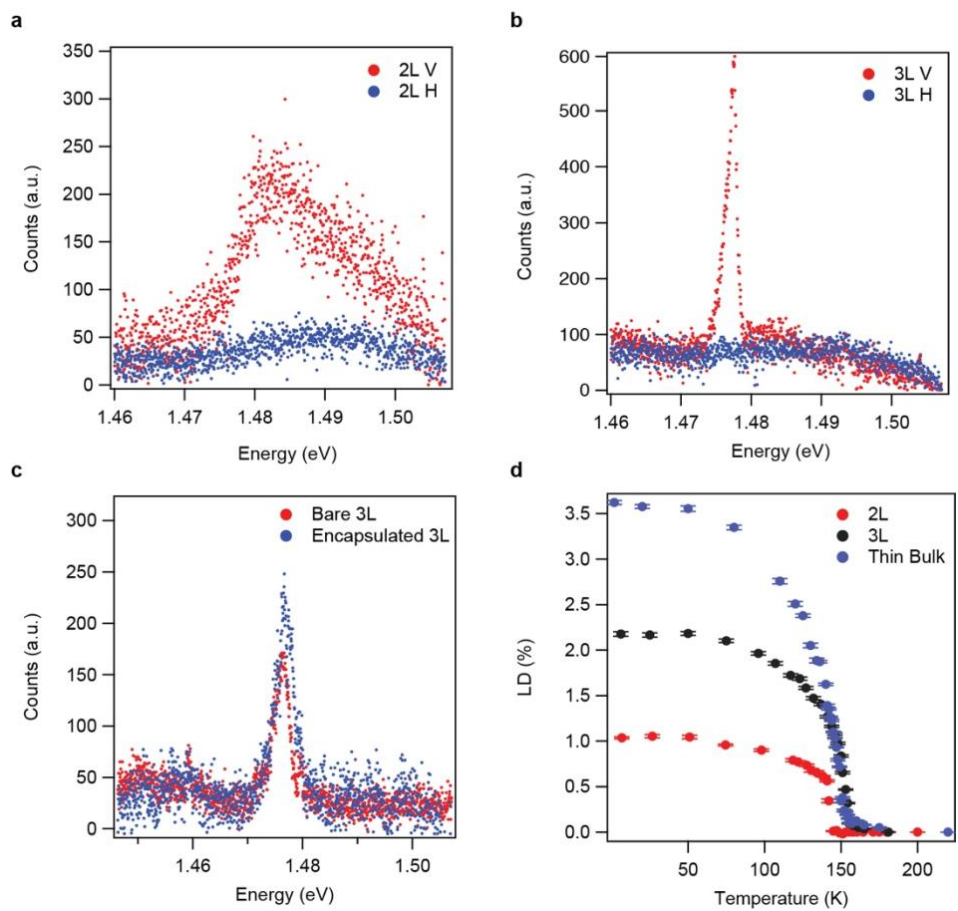

**Extended Data Figure 6 | Additional measurements of atomically thin samples.** Polarization resolved photoluminescence spectra of **a**, bilayer and **b,** trilayer at 15 K. **c**, Comparison of the photoluminescence spectra with and without hBN encapsulation of a trilayer. There is no appreciable linewidth difference between the two. **d**, Temperature dependent linear dichroism of bilayer, trilayer, and thin bulk crystal. Error bars were obtained from a sinusoidal fit of the rotational LD measurements.

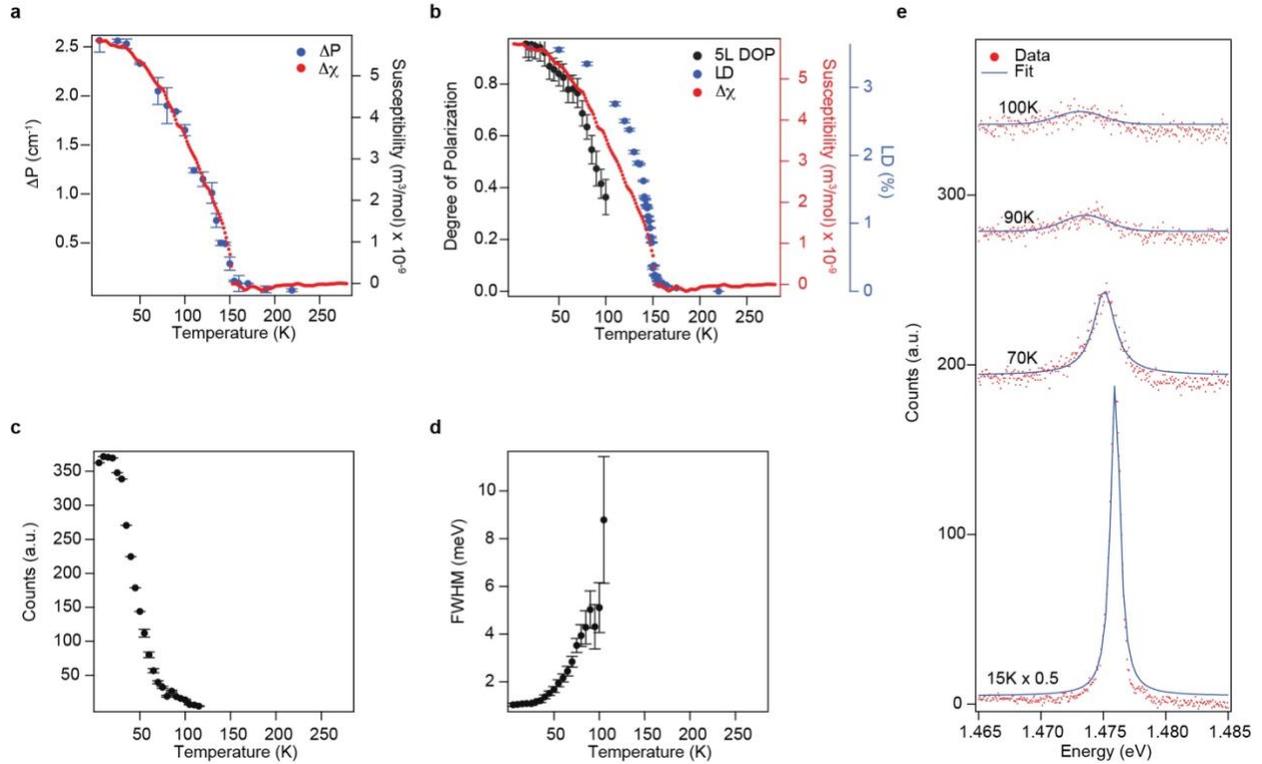

**Extended Data Figure 7 | Comparison of temperature dependent optical properties of a five-layer sample with in-plane magnetic susceptibility anisotropy Δχ of bulk crystal NiPS$_3$. a**, Temperature dependent energy splitting ΔP of the 180 cm$^{-1}$ Raman mode. Ref. 26 shows that ΔP can be used to probe the zigzag antiferromagnetic order. Here, we compare ΔP with Δχ. Remarkably, both physical quantities share similar temperature dependent behavior. This comparison shows that ΔP originates from Δχ, and Δχ in five-layer sample is similar to that of the bulk crystal. Error bars were obtained from a Lorentzian fit of the Raman peaks. **b,** Temperature dependent degree of linear polarization (DOP) and linear dichroism (LD) overlaid with Δχ. Both DOP and LD resemble Δχ, although DOP are limited below 100 K due to the vanishing photoluminescence above 100 K. **c**, Temperature dependent photoluminescence intensity and **d**, full width at half maximum (FWHM). Error bars in figures **c** and **d** represent the confidence bounds of a Lorentzian fit. **e,** Photoluminescence spectra and their respective Lorentzian fits at select temperatures.

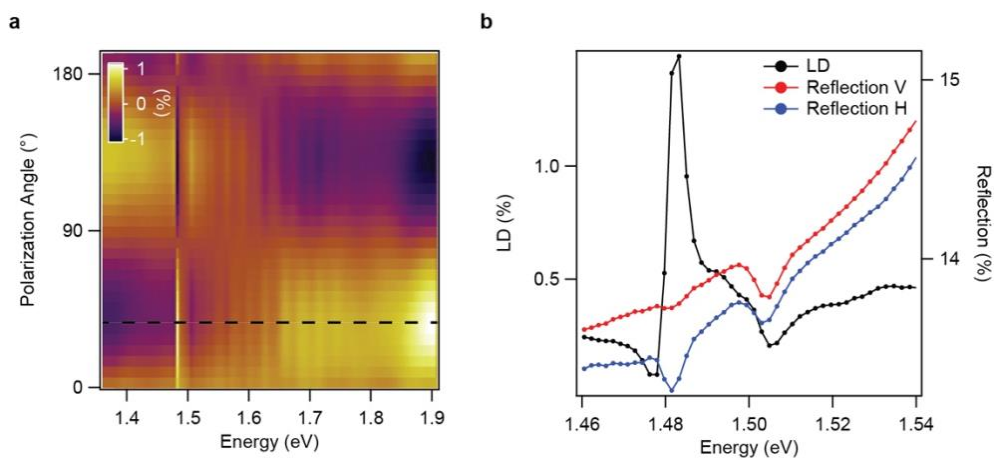

**Extended Data Figure 8 | Polarization-resolved optical reflection and its connection to linear dichroism spectrum. a**, Polarization dependent linear dichroism (LD) spectra. Black dashed line corresponds to LD presented in Fig. 4a (main text), which is obtained by the excitation along the vertical axis. **b**, Red and blue curves are vertically and horizontally polarized optical reflection spectra, overlaid with the LD spectrum (black curve). Two resonances are observed[36]. The low energy resonance near 1.4815 eV has much stronger polarization dependence than the one near 1.504 eV. The difference of these two peaks partially contribute to the observed LD lineshape.

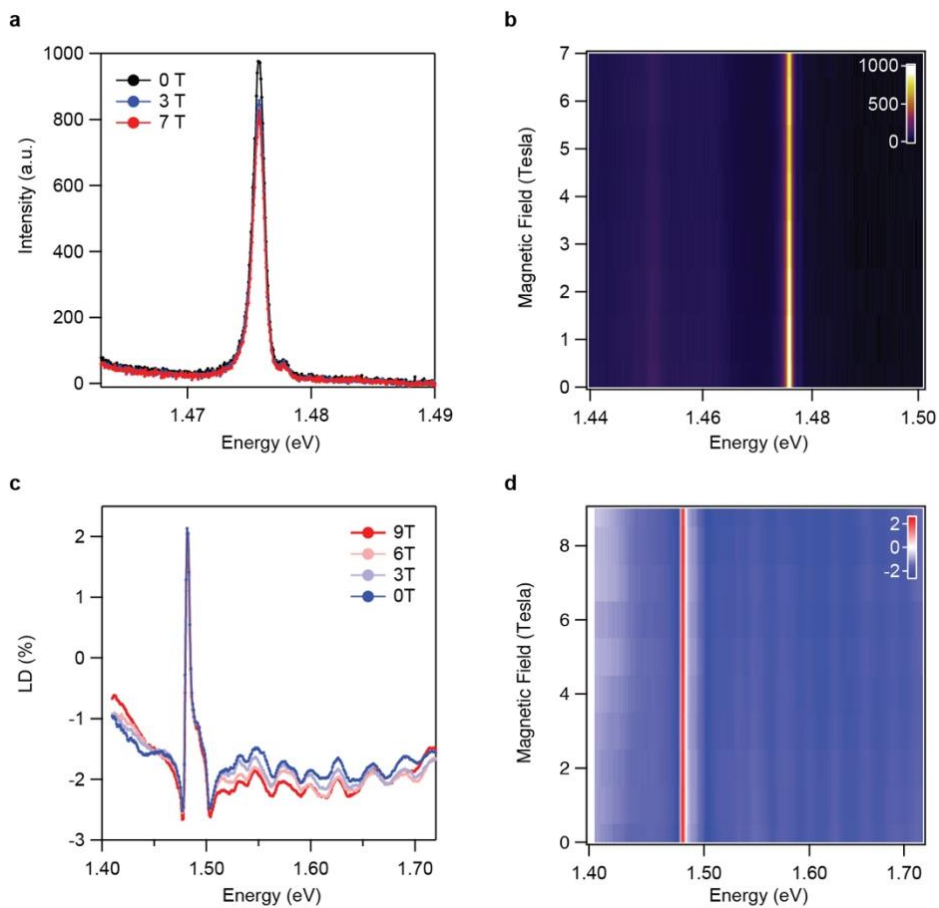

**Extended Data Figure 9 | Magnetic-field dependent photoluminescence and linear dichroism. a,** Photoluminescence spectra of a five-layer sample at selected magnetic fields. **b,** Photoluminescence intensity plot of the same 5L sample as a function of magnetic field and photon energy. **c,** LD of a bulk crystal at selected magnetic fields. **d**, LD intensity plot as a function of magnetic fields and photon energy. All experiment data are taken at 15 K.